\documentclass[%
 aps,
 prb,
 jmp,%
 amsmath,amssymb,
 reprint,%
]{revtex4-1}

\usepackage[english]{babel}

\usepackage{graphicx}
\usepackage{dcolumn}
\usepackage{bm}
\usepackage{epstopdf}

\begin{document}

\title[Micromagnetic view on ultrafast magnon generation by femtosecond spin current pulses]{Micromagnetic view on ultrafast magnon generation by femtosecond spin current pulses}

\date{\today}

\author{Henning Ulrichs}
\email{hulrich@gwdg.de}
\affiliation{ 
I. Physical Institute, Georg-August University of G\"ottingen, Friedrich-Hund-Platz 1, 37077 G\"ottingen, Germany%
}%

\author{Ilya Razdolski}
\affiliation{ 
Physical Chemistry Department, Fritz Haber Institute of the Max Planck Society, Faradayweg 4-6, 14195 Berlin, Germany%
}%

\begin{abstract}
In this Article we discuss a micromagnetic modelling approach to describe the ultrafast spin-transfer torque excitation of coherent and incoherent magnons on the nanoscale. Implementing the action of a femtosecond spin current pulse entering an orthogonally magnetized thin ferromagnetic film, we reproduce recent experimental results and reveal the factors responsible for the unequal excitation efficiency of various spin waves.
Our findings are in an excellent agreement with the results of an analytical description of spin-wave excitation based on classical kinetic equations.
Furthermore, we suggest an experimental design allowing for the excitation of laterally propagating spin waves beyond the optical diffraction limit. Our findings demonstrate that the classical micromagnetic picture retains its predictive and interpretative power on femtosecond temporal and nanometer spatial scales.

\end{abstract}

\keywords{Spin currents, spin transfer torque, magnon generation, spin waves, spin dynamics, micromagnetic simulations}
  
\maketitle

\section{Introduction}

Stimulated by the seminal experiment by Beaurepaire et al. \cite{Beaurepaire1996} and the quest for ultrafast opto-magnetic recording, an immense amount of knowledge on the incoherent laser-induced spin dynamics (i.e. ultrafast demagnetization) in a large variety of materials has been accumulated over the years \cite{KKR2010}. Simultaneously, femtosecond optical excitation of coherent spin dynamics
was discovered \cite{Kikkawa1998,Bigot2009}, encompassing an extremely broad range of timescales which are governed by the intrinsic properties of magnetic systems. The temporal limitation for the excited spin modes is often pertinent to the pulse duration of the light source, typically on the order of 10-100 fs. However, the spectrum of the accessible inhomogeneous (with nonzero wavevector $k$) spin wave modes is governed by the spatial inhomogeneity scale of the excitation. In other words, on top of the temporal requirements, non-uniform spin wave modes can only be generated if their wavevectors $k$ are contained in the spectrum of the spatially inhomogeneous stimulus. For the excitation with visible (VIS) or near-infrared (NIR) light, the optical penetration depth $\delta\approx10-15$ nm serves as a good estimation for the excitation limit of perpendicular spin waves in metallic media \cite{vanKampen2002}. Optical excitation of the in-plane propagating spin waves is even more restrictive, as the allowed $k$ values are governed by the diffraction-limited beam spot size ($\gtrsim 10^3-10^4$~nm) \cite{SatoNatPhot2012, AuPRL2013}.

Yet, in recent experiments, a strong interfacial localization of the spin transfer torque exerted by spin polarized currents enabled the excitation of spin waves with much larger wavevectors \cite{Razdolski2017,Lalieu2017}. In particular, perpendicular standing spin waves (PSSW) with $f=0.55$~THz and $k\sim1$~nm$^{-1}$ have been detected \cite{Razdolski2017}. The wavelengths of these excitations approach the exchange length $l_{\rm ex}$ (a few nm in Fe \cite{Frei1957}), where macroscopic spin models are likely to break down \cite{Berkov2008}. Enabling the expansion of ultrafast photo-magnonics \cite{Djordjevic2007,Lenk2011} onto the nanometer scale, these findings simultaneously question the applicability of conventional modeling of spin dynamics in these extreme conditions. 

For conventional magnonics, micromagnetic simulation is an indispensable tool for both prediction, and interpretation of static and dynamic magnetic properties \cite{Hertel2004,Kruglyak2010,Kim2010,Iwasaki2013,Dvornik2013,Ulrichs2014,Banerjee2017}. 
In this article, we show that the micromagnetic modelling approach is also suitable for ultrafast processes on nanometer scales. In particular, we set up a micromagnetic model to first reproduce the recent experimental findings \cite{Razdolski2017} of ultrafast optical magnon generation. In this study, Fe/Au/Fe trilayers were optically pumped from one side, generating a spin current pulse which traverses the Au spacer and then interacts with the second Fe layer, resulting in the excitation of high-frequency spin dynamics in the THz domain.
Here we develop a micromagnetic model featuring ultrafast spin-transfer torque perturbation and verify that it can accurately reproduce the experimentally observed spin dynamics. We further identify important factors governing the excitation efficiency and energy transfer into the PSSW modes in thin ferromagnetic films. Complementing a recent theoretical work \cite{Balaz2017} on laser-generated superdiffusive spin transport in non-collinear spin valve systems and resulting macrospin dynamics in ferromagnets, our results open the door to understanding spin current-driven magnetism on the nanoscale.
Later on, we include thermally activated, incoherent magnetic fluctuations in the model, and show that ultrafast spin currents can effectively cool or heat such thermal magnon ensembles, in agreement with experimental observations on slower timescales \cite{PhysRevLett.107.107204}. In the outlook, we outline the topological possibilities for the spin current-mediated generation of in-plane propagating, large-$k$ spin waves beyond the optical diffraction limit.

\begin{figure}
\includegraphics[width=8.5cm]{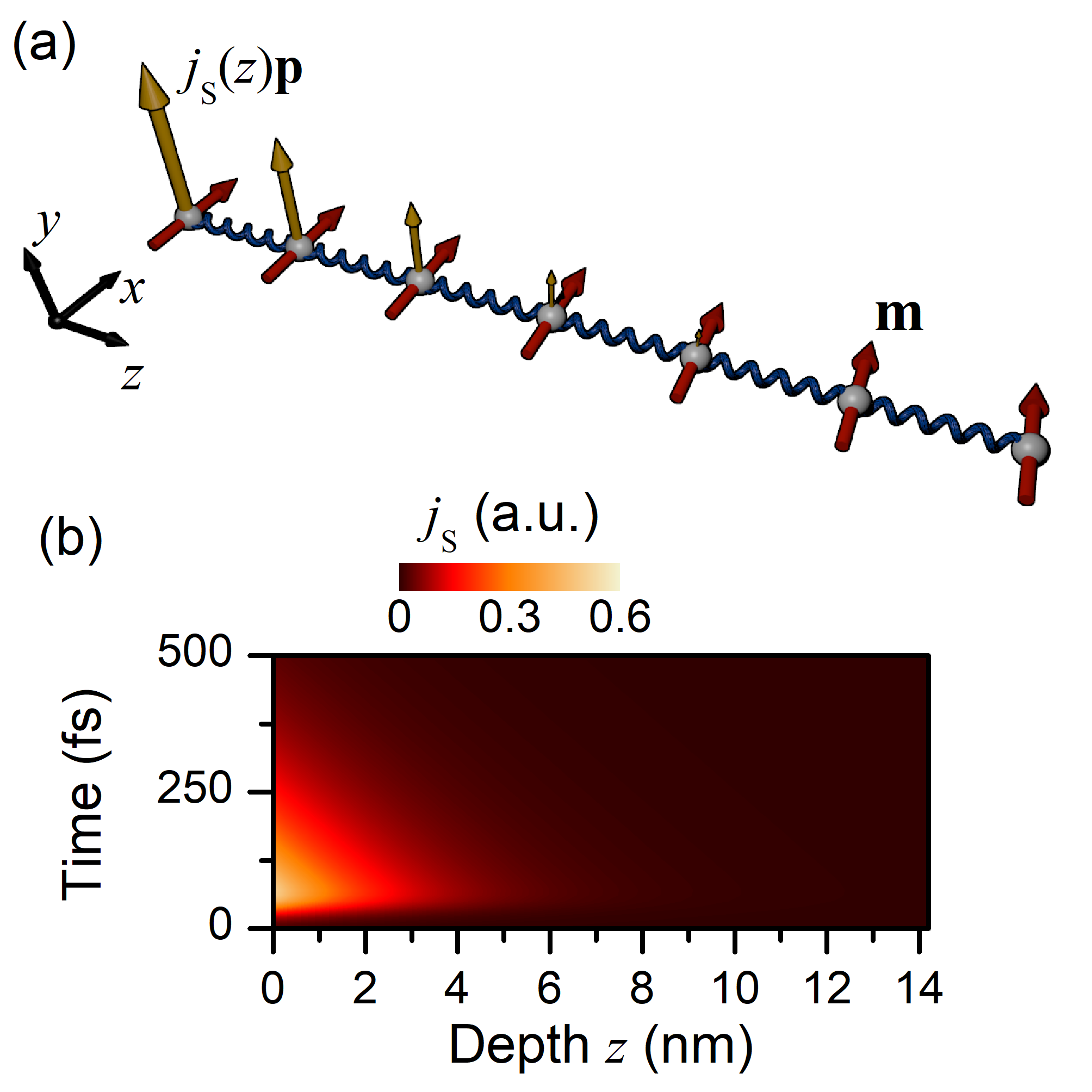}
\caption{Model details. (a) Sketch of the micromagnetic model. Besides incorporating the spin current, the model optionally includes thermal fluctuations. (b) Spatio-temporal dependence of the spin current entering the film, assuming $\lambda_{\rm STT}=2\,$nm.}
\label{fig:figure1}
\end{figure}

\section{Model details}
Our model is visualized in Figure \ref{fig:figure1}(a). It consists of a single ferromagnetic layer of Fe, with a thickness of $d=14.2\,$nm. We take as material parameters a saturation magnetization of $\mu_0M_0=2.1\,$T, an exchange constant of $A=19\,$pJ/m (corresponding to the  exchange stiffness $D=280$~${\AA}^2\cdot$~meV in Fe from Ref.~\cite{MookNicklow}), and a uniaxial anisotropy along $\mathbf x$ with strength $K_u=45956\,$J/m$^3$, and assume a Gilbert damping factor of $\alpha=0.008$. 

For the demonstration of coherent magnon generation, we simulate a cube of size $1.16\times 1.16 \times 14.2\,$nm$^3$ subdivided into $N_x \times N_y \times N_z=2 \times 2 \times 24$ cells. Periodic boundary conditions in $\mathbf x$ and $\mathbf y$ direction enlarge this cube into an infinitely extended film.  For the demonstration of incoherent magnon creation and annihilation, we simulate a larger cube of size $150 \times 150 \times 14.2\,$nm$^3$ subdivided into $N_x \times N_y \times N_z=256 \times 256 \times 24$ cells. Note that in this case, an additional magnetic field representing thermal fluctuations is switched on. The spin current $j_s$ with polarization $\mathbf p$ enters the system at $z=0$, and has the following empirical spatio-temporal form:

\begin{equation}
j_s=\frac{\hbar}{2e}j_0 e^{-z/\lambda_{\rm STT}}\frac{e^{-t/\tau_2}}{1+e^{-(t-t_0)/\tau_1}}.
\label{eq:js}
\end{equation}

Here the penetration depth of the spin current is $\lambda_{\rm STT}=2\,$nm, as estimated in Ref.~\cite{Razdolski2017}, and the temporal profile of the spin current pulse (Fig.~\ref{fig:figure1},b) is approximated with an analytic function \eqref{eq:js} with $t_0=50\,$fs, $\tau_1=10\,$fs, and $\tau_2=150\,$fs, closely reproducing the results of Ref.~\cite{Alekhin2017}. We use the software package mumax$^3$ (Ref.~\cite{vansteenkiste2014}) to model the effect of a spin current pulse on the local magnetization inside the Fe film by augmenting the Landau-Lifshitz-Gilbert equation with the spin transfer torque (STT) term $\mathbf{\tau}_{\rm STT}$ proposed by Slonczewski \cite{SLONCZEWSKI1996L1}:

\begin{equation}
\mathbf{\tau}_{\rm STT}=\gamma\frac{N_z}{d \mu_0 M_0} j_s \mathbf M \times \mathbf M \times \mathbf p. \label{eq:STT}
\end{equation}

\begin{figure}[h!]
\includegraphics[width=8.25cm]{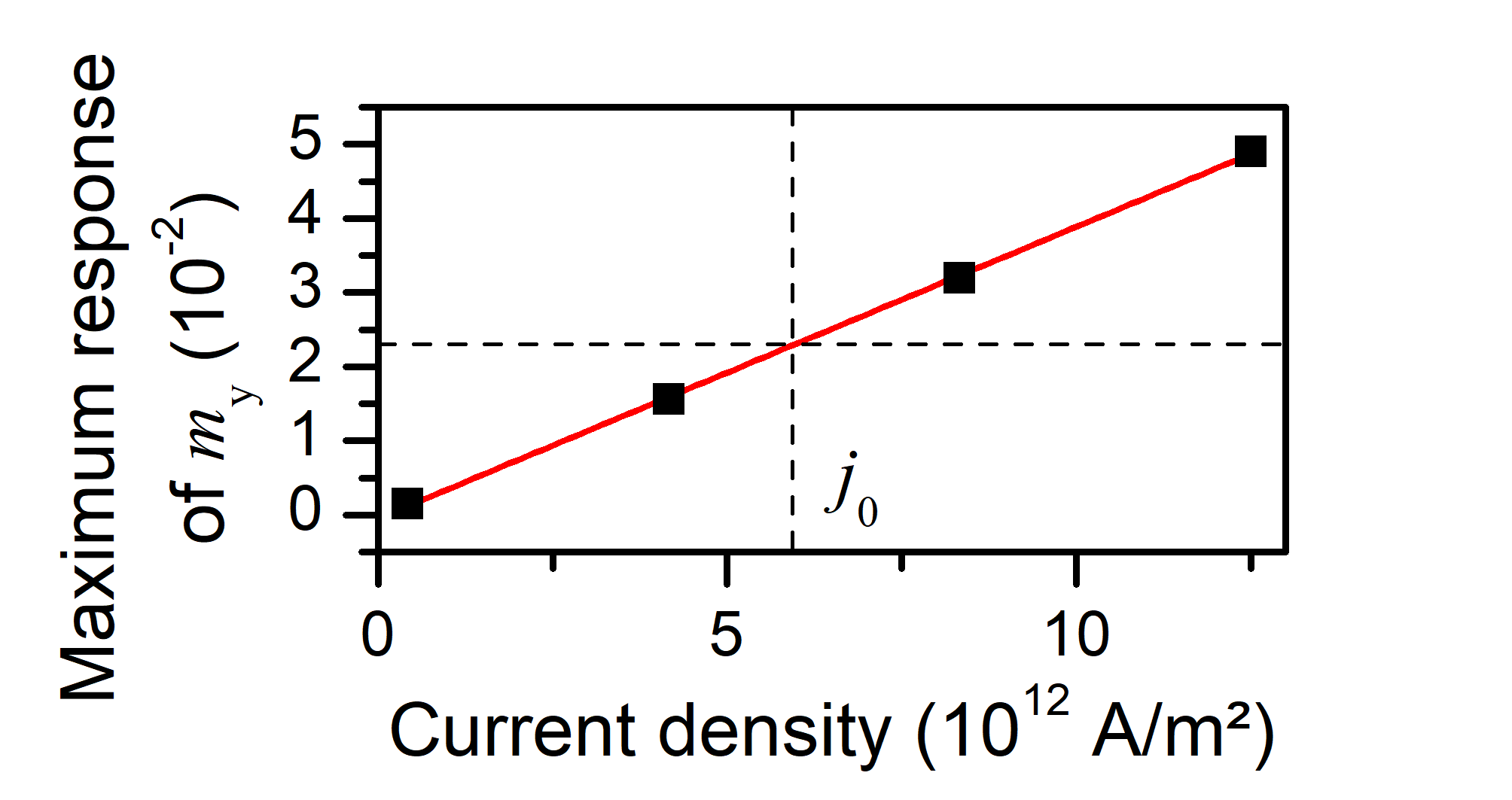}
\caption{Determination of $j_0$ from the dependence of the maximum of $m_y$ on the current density $j$. The horizontal dashed line marks the $m_y$ experimental value which corresponds to $j_0=5.9\cdot 10^{12}\,$A/m$^2$ (see the vertical dashed line).}
\label{fig:figureS2}
\end{figure}

\noindent The reported in-plane excursion of the magnetization of $m_y=\frac{M_y}{M_0}=0.023$ directly after the spin current pulse arrival allows us to determine the respective current density to be used in the simulations. For this purpose,  we 
systematically varied the current density and analyzed the temporal evolution of the in-plane component $m_y$. The maximum excursion appears shortly after the spin current pulse maximum, which is in agreement with Ref.~\cite{Razdolski2017} (see Figure 4(b) therein). Figure \ref{fig:figureS2} shows that the maximum depends linearly on the applied current density. The linear interpolation intersects with the horizontal dashed line defined by the experimental value of $m_y$ at $j_0=5.9\cdot 10^{12}$A/m$^2$. This result in an excellent agreement with the one obtained from the spin transfer density ($7~\mu_B$/nm$^2$) evaluated in Ref.~\cite{Razdolski2017} ($\approx6\cdot10^{12}\,$A/m$^2$), thus reinforcing our micromagnetic model.

\noindent Further, this $j_0$ value is realistic, as it corresponds to the $\sim 10\%$  spin transport-induced demagnetization of a $10$~nm-thick Fe film within $200$~fs, in agreement with the latest results obtained within the superdiffusive transport model \cite{Balaz2017}. Moreover, it is close to the values reported in other works ($10^{12}-10^{13}\,$A/m$^2$, Ref.~\cite{MelnikovPRL11}, and $10^{13}\,$A/m$^2$, Ref.~\cite{BattiatoPRL16}). Similar numbers can be further obtained from the work of Choi et al. \cite{ChoiPRB14} ($\sim 10^{12}\,$A/m$^2$).

\begin{figure*}
\includegraphics[width=17.5cm]{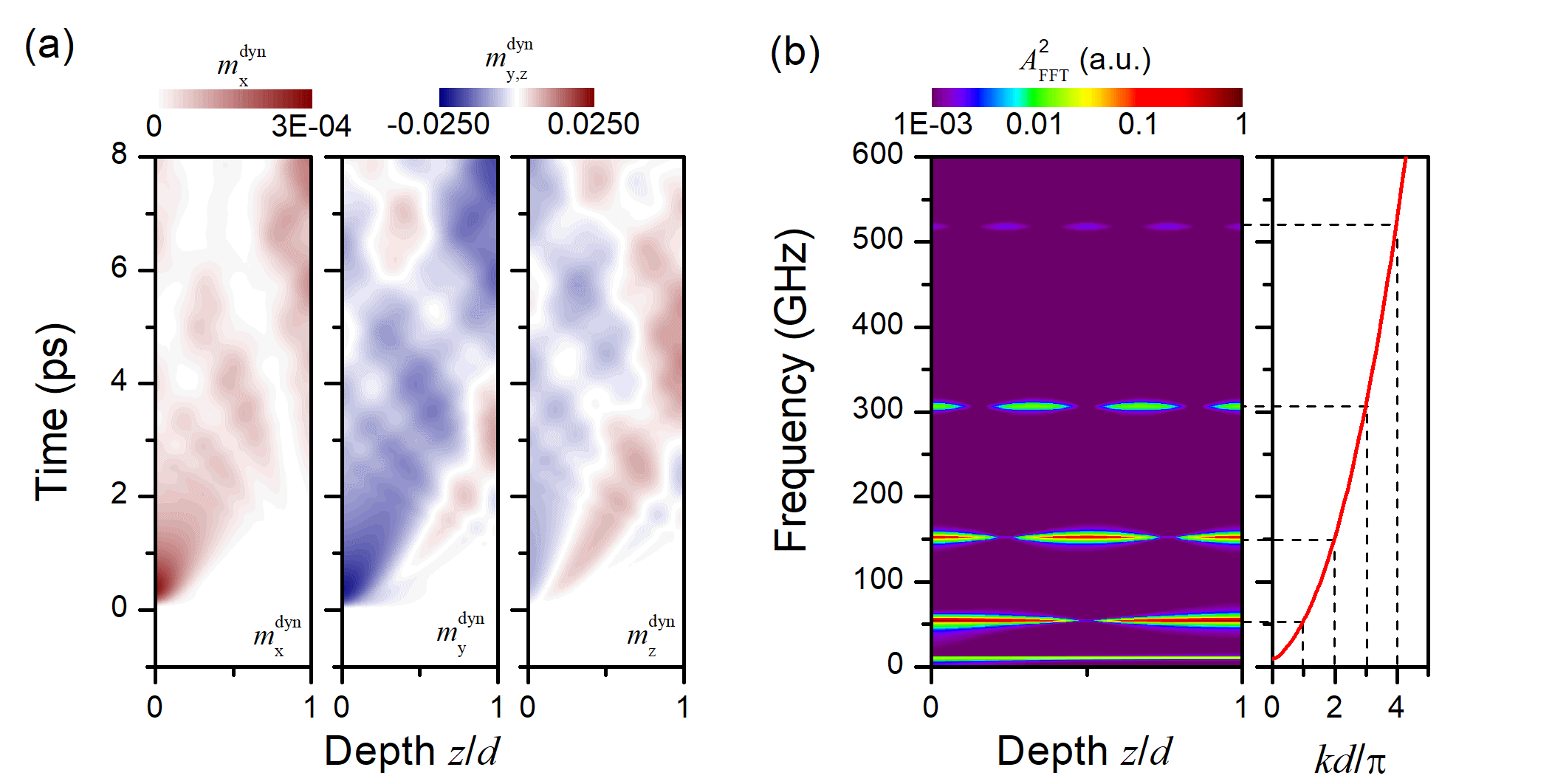}
\caption{Magnetization dynamics driven by the ultrafast spin-transfer torque. (a) Spatio-temporal plot of the dynamic part $\langle m_i^{\rm dyn}\rangle$ of laterally averaged dynamic magnetization components. The left panel of (b) shows a spatial Fourier transform of the $\langle m_z\rangle$ data shown in the right panel of (a), and the right panel of (b) shows the spin-wave dispersion $f(k)$ from Eq.~(3). The dashed lines relate the numerical response to the analytic theory.}
\label{fig:figure2}
\end{figure*}

\section{Results}
\subsection{Coherent magnon generation}

In the first numerical experiment, we prepared a spin current pulse with polarization $\mathbf p \parallel \mathbf y \perp \mathbf m $. Then, according to Eq.~\eqref{eq:STT} the spin torque will be $\mathbf \tau_{\rm STT} \perp \mathbf m$. The subsequent spatially-resolved spin dynamics was computed for a total time of $1\,$ns after the spin current pulse peak. Figure \ref{fig:figure2}(a) shows $m_i^{\rm dyn}$, which is the dynamic part of the laterally averaged magnetization component $\langle m_i\rangle_{x,y}(z,t)$ ($i=x,\,y,\,z$) in each layer of the Fe film for the first $8\,$ps. One can see how the spin current induces the formation of a localized wave packet which then expands. Note that the quadratic dispersion of exchange-dominated spin waves \cite{0022-3719-19-35-014}

\begin{equation}
	f(k)=\frac{\gamma\mu_0}{2\pi}\sqrt{\left(H_{\rm an}+\frac{2A}{M_0}k^2\right)\cdot\left(H_{\rm an}+\frac{2A}{M_0}k^2+ M_0\right)}\label{eq:dispersion}
\end{equation}

\noindent is responsible for the quick spatial broadening of the spin-wave packet. Here, $\frac{\gamma}{2\pi}\approx28$~GHz/T is the gyromagnetic ratio, $H_{\rm an}=2K_u/M_0$ is the in-plane crystalline anisotropy field. The front of the wave packet travels with a characteristic speed of about $7\,$ nm/ps corresponding to the group velocity of the magnons with the largest wavevectors contained in the excitation spectrum. After about $2\,$ps, the pulse has reached the surface of the Fe film. The spin dynamics for times $t>2\,$ps is formed by a complex interference pattern which can be well analyzed by a Fourier transformation in time, as shown in Figure \ref{fig:figure2}(b). There, the local squared Fourier amplitude $A^2_{\rm FFT}\lbrace\langle m_z\rangle_{x,y}\rbrace(z,f)$ is depicted. In agreement with the experiment \cite{Razdolski2017}, this representation reveals that the pattern from Figure \ref{fig:figure2}(a) can be understood as a superposition of standing spin waves. This is emphasized by the right panel in Figure \ref{fig:figure2}(b), which shows the dispersion (\ref{eq:dispersion}), plotted as a function of a dimensionless wave number $\kappa=kd/\pi$. Prominent spin dynamics can be found at integer $\kappa_n=k_nd/\pi=0,\,1,\,2,\,\ldots$, and the corresponding eigenmode frequencies $f_n=f(\kappa_n)$.

\begin{figure*}
\includegraphics[width=17.5cm]{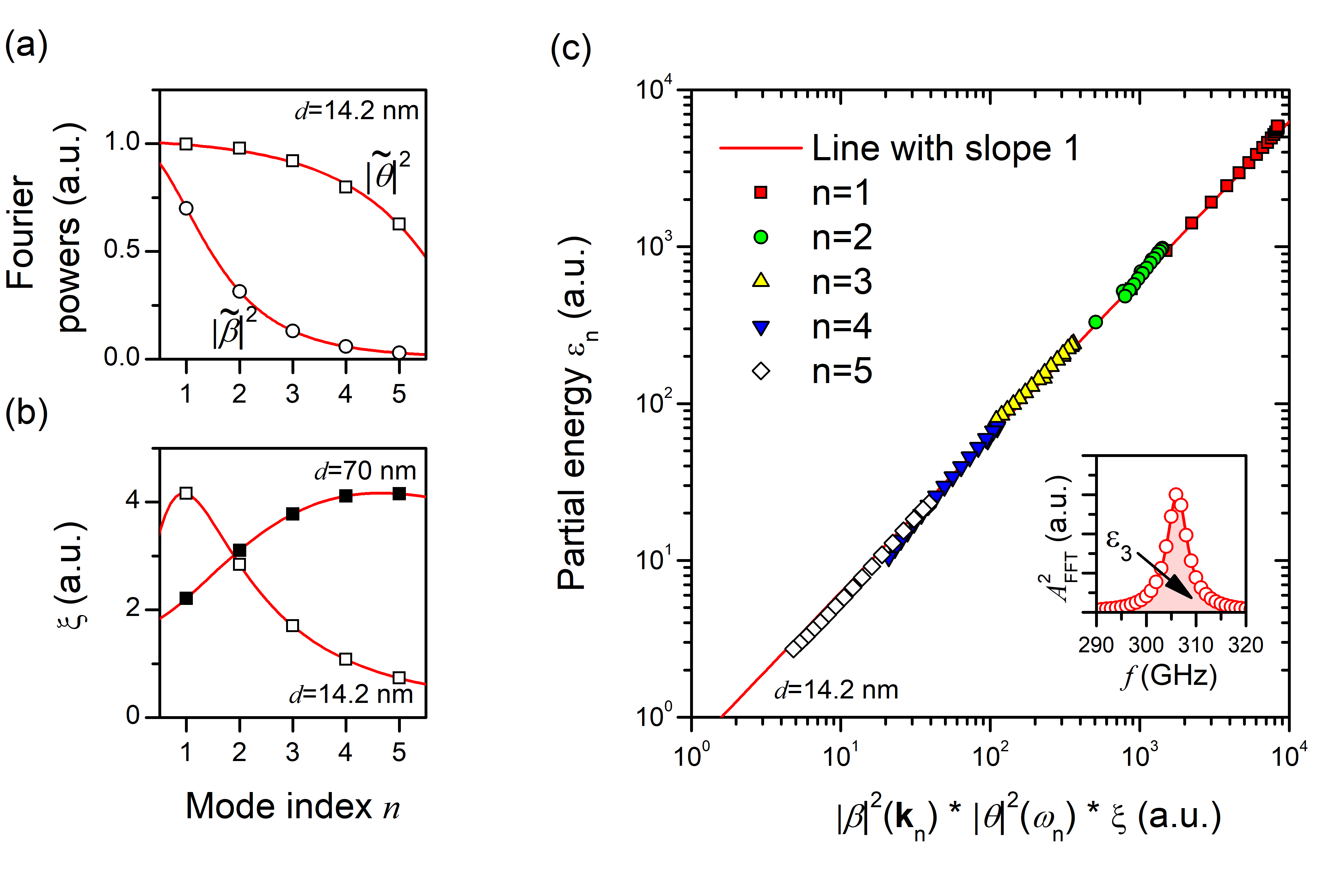}
\caption{PSSW excitation efficiency. (a) Squared spatial $|\tilde\beta|^2$ and temporal $|\tilde\theta|^2$ Fourier amplitudes of the spin-current pulse \eqref{eq:js} as a function of the PSSW mode index for $d=14.2\,$nm and $\lambda_{\rm STT}=2\,$nm. The lines are shown as a guide to the eye. (b) Additional $\xi$ factor as a function of the PSSW mode index for two different Fe film thicknesses $d$ (14.2 nm, open squares, and 70 nm, full squares). (c) Correlation-like double logarithmic plot of the partial energy $\varepsilon_n$ versus the product of the various factors which govern the energy transfer, $|\tilde\beta|^2\cdot |\tilde\theta|^2\cdot\xi$. The different colors indicate various PSSW eigenmodes while within a single color, the data points are obtained for different values of $\lambda_{\rm STT}$. The solid line is a linear fit to the data.}
\label{fig:figure3}
\end{figure*}

Besides reproducing experimental findings, the micromagnetic model allows for an identification of the parameters that govern the mode-specific excitation efficiency. As such, we will now discuss the energy transfer into the different spin wave modes. For this purpose, we analyze in detail squared FFT amplitude spectra at the interface of the Fe film, $z=0$. This choice is motivated by the first order boundary conditions for the spin wave modes ensuring the largest amplitude of the oscillations of the spin projections at the interfaces of the Fe film \cite{0022-3719-19-35-014}. The integration of the $A^2_{\rm FFT}(f)$ spectrum in the vicinity of the peaks corresponding to the excitation of the different spin-wave modes yields the partial energies $\varepsilon_n$ associated with each of the modes:

\begin{equation}
\varepsilon_n=\int_{f_n-\Delta f}^{f_n+\Delta f} A^2_{\rm FFT}(f)~df    
\end{equation}

\noindent This quantity is proportional to the energy transferred into this mode. In what follows, we shall focus on various factors which contribute to $\varepsilon_n$, aiming at understanding their importance for a comparative analysis of the excitation efficiency of the PSSW modes in thin ferromagnetic films. 

In the Supplementary Information we develop an analytic formalism based on Holstein-Primakoff transformations, which is capable of deriving the contributing factors in detail. Importantly, the energy supplied by a spin current pulse $j_s(z,t)$ is proportional to the product $|\tilde\beta|^2 (\mathbf k_n) \cdot |\tilde\theta|^2 \left(\omega_n\right)$ of the spatial and temporal Fourier powers of $j_s(z,t)$, evaluated at $k=k_n$, and $\omega=\omega_n$. These two factors are shown in Fig.~\ref{fig:figure3}(a) for $\lambda_{\rm STT}=2\,$nm. Note that the temporal factor $|\tilde\theta|^2$ is only important when the oscillation period $T_n=1/f_n$ approaches the duration of the spin current stimulus $\sim\tau_2$. Thus, in our case for $n<5$ and $\sim 250$~fs spin current pulse duration, the spatial factor $|\tilde\beta|^2$ plays a dominant role in determining the relative excitation efficiency of the PSSW modes. 

Further, we identify the material parameter-dependent susceptibility of the different spin-wave modes which, together with their ellipticity $\mathcal{E}_\mathbf k$, gives rise to another factor $\xi=\frac{\Gamma_\mathbf k}{\omega_\mathbf k^2}\frac{1-\mathcal{E}_\mathbf k}{2-\mathcal{E}_\mathbf k}$ (see Supplementary Information for details). Figure \ref{fig:figure3}(b) shows the dependence of $\xi$ on the mode number $n$ for two different film thicknesses. For thin films exemplified here as $d=14.2\,$nm, the $\xi$ factor peaks at $n=1$ and further decays for higher modes. As this dependence is similar to the behaviour of $\beta_{\mathbf k_n}$, the role of $\xi$ for the relative excitation efficiency consists in emphasizing the mode with $n=1$, consistent with the experimental data \cite{Razdolski2017}. However, for thicker films, the maximum of $\xi$ is shifted towards higher modes ($n=5$ for $d=70\,$nm). Note that because in those films both $\omega$ and $k$ only slightly increase with $n$, for small $n$ (when $k_n\lambda_{\rm STT}\ll 1$, $\omega_n\tau_2\ll 1$) the other factors $|\tilde\beta|^2 (\mathbf k_n)$ and $|\tilde\theta|^2 \left(\omega_n\right)$ are both almost constant. As such, up to much larger $n$, the dynamical response is dominated by $\xi$ and thus can be enhanced at higher ($n>1$) order spin wave modes. Supported by the analytic theory, we expect in summary a linear relation

\begin{equation}
	\varepsilon_n\propto |\tilde\beta|^2 (\mathbf k_n) \cdot |\tilde\theta|^2 \left(\omega_n\right)\cdot \xi. \label{eq:factors} 
\end{equation}

\noindent Indeed, Figure \ref{fig:figure3}(c) shows that a linear scaling law holds over four decades. The data shown here were obtained by varying $\lambda_{\rm STT}$ between $0.5\,$nm and $5\,$nm. The excellent agreement between the predictions of the analytic calculations and the results of numerical simulations emphasizes that all significant factors are accounted for in Equation \eqref{eq:factors}.

\begin{figure}
\includegraphics[width=8.25cm]{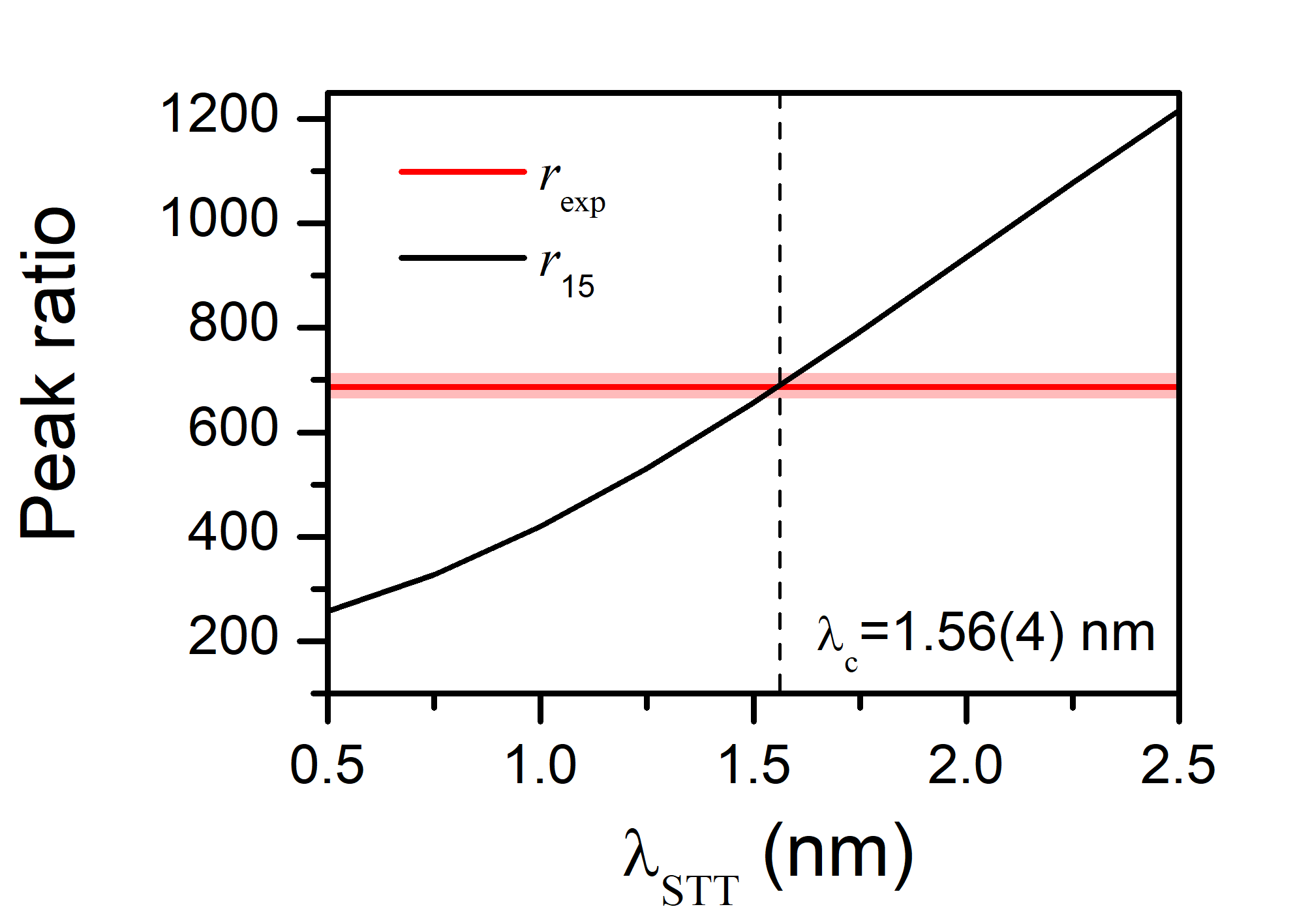}
\caption{Determination of a lower boundary for the spin transfer torque characteristic depth $\lambda_{\rm STT}$. Simulated (black) and experimental (red) peak ratios as a function of $\lambda_{\rm STT}$. The red shaded region indicates the error bar around the data from the experiments. The vertical dashed line shows the estimated lower boundary $\lambda_c$.}
\label{fig:figurelambdastt}
\end{figure}

Having outlined the factors contributing to the excitation efficiency of the PSSW modes in thin films, we can now build a bridge to the experiment. We note that in Ref.~\cite{Razdolski2017} an upper boundary for $\lambda_{\rm STT}$ has been identified, based on the spin current pulse ability to excite the spin wave mode with $n=4$. Here, the above mentioned formalism enables the determination of a {\it lower} $\lambda_{\rm STT}$ boundary. The critical conditions for that rely on the fact that the excitation efficiency for $n=5$ mode was found insufficient for its unambiguous detection in the experimental data. Clearly, for smaller $\lambda_{\rm STT}$ the $n=5$ mode will be more strongly excited. As such, we can calculate the ratio $r_{15}=s_{15}\cdot A_1/A_5$ for various $\lambda_{\rm STT}$, where $A_n$ is the Fourier amplitude of the $n$-th PSSW mode at the interface. The correction factor $s_{15}$ takes into account that in the experiment the MOKE in-depth sensitity function $w(z)$ is responsible for the fact that different PSSW modes contribute unequally to the total MOKE signal. We calculate  $w(z)$ using an optical transfer matrix method \cite{Zak90}, and determine $s_{15}$, in order to enable direct comparison with the experimental peak-to-noise ratio $r_{\rm exp}=\frac{A_1^{\rm exp}}{A_N}$. We arrive at the following condition for $\lambda_{\rm STT}$:

\begin{equation}
   r_{15}(\lambda_{\rm STT})\geqslant\frac{A_1^{\rm exp}}{A_N},
   \label{eq:lambdaboundary}
\end{equation}

\noindent In Figure \ref{fig:figurelambdastt} we show the dependence of the peak ratios on $\lambda_{\rm STT}$. The intersection of the red (experimental) and black (simulated) curves indicates the lower boundary for $\lambda_{\rm STT}$ of about $1.56$~nm.

\subsection{Incoherent magnon creation and annihilation}

\begin{figure}
\includegraphics[width=8.25cm]{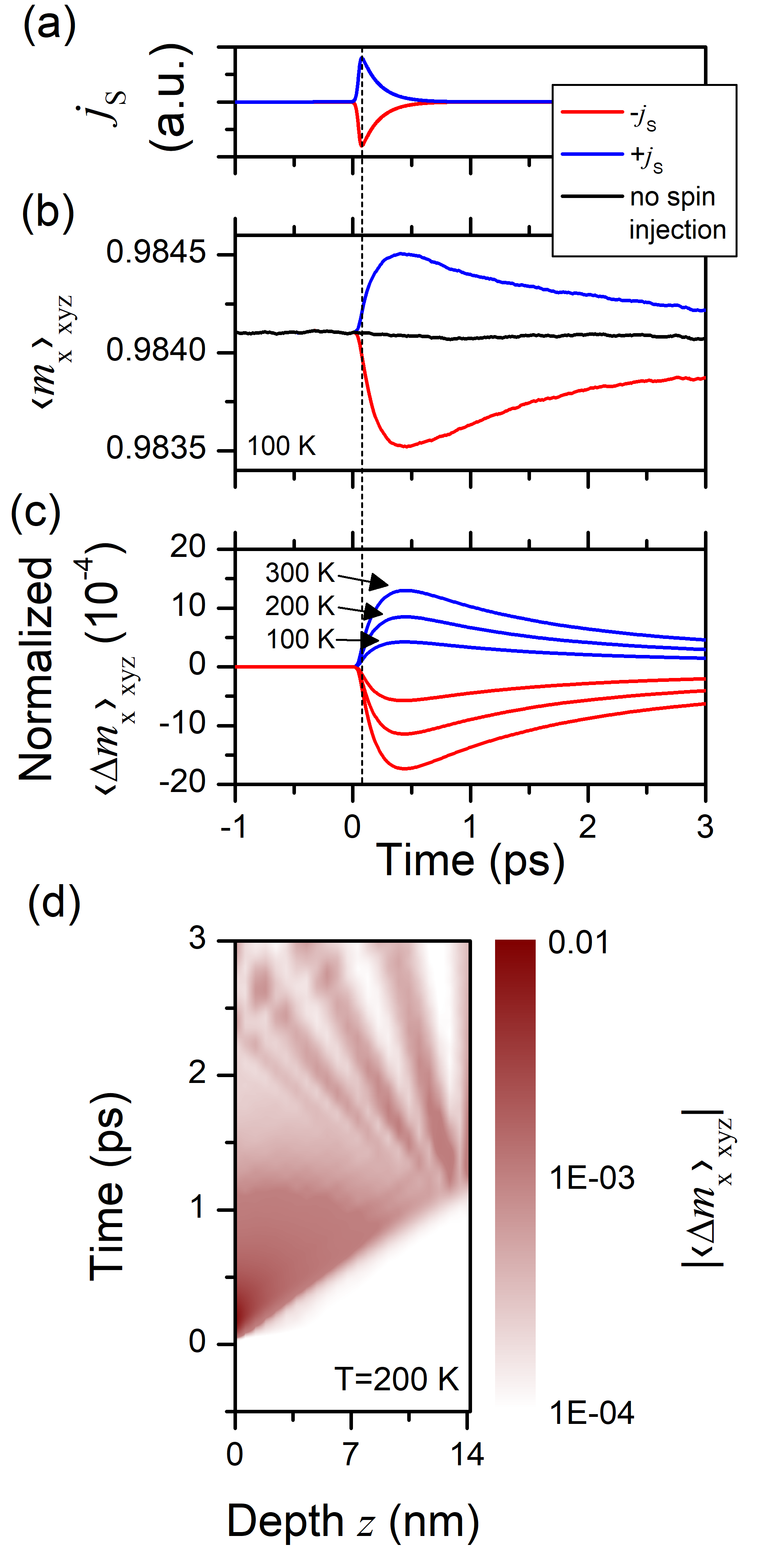}
\caption{Suppression and enhancement of incoherent magnon dynamics by spin currents. (a) Temporal shape of the spin-current pulse $j_s(t)$ of the two opposite polarities. (b) Simulated time dependence of the spatially averaged longitudinal magnetization component $\langle m_x\rangle$ at $T=100\,$K. (c) Variations of spatially averaged longitudinal magnetization component $\langle m_x\rangle$, normalized to that obtained in the case of no spin current injection. Multiple lines show the results calculated for three various temperatures, as indicated by dark arrows. The dashed line in (a) through (c) indicates the peak time of the spin current pulse. (d) Layer-resolved, only laterally averaged, normalized change of the longitudinal magnetization component $\langle m_x\rangle$ obtained at the temperature of 200 K.}
\label{fig:figure4}
\end{figure}

As a second numerical experiment, we prepared a spin current pulse whose polarization $\mathbf p$ is collinear to $\mathbf x$, and thus to the Fe magnetization $\mathbf m$. Then, according to Equation (\ref{eq:STT}) at temperature $T=0$ the spin torque $\tau_{\rm STT}$ vanishes, and no coherent spin dynamics can be expected. At finite temperatures this expectation is, however, misleading due to the magnetic fluctuations present in the ferromagnet. Practically, these fluctuations manifest as a reduction of the average magnetization. Averaging in time gives the transverse components $\langle m_{y,z} \rangle_t=0$, and the longitudinal component $\langle m_{x} \rangle_t<1$. The larger the temperature, the smaller is the longitudinal magnetization component. Switching on a spin current pulse with $\mathbf p$ collinear to $\mathbf x$ acts only on the transverse components, which are momentarily nonzero. It is well known from conventional magnonics, that the resulting torque is damping- or antidamping-like, and that thermal fluctuation will therefore be suppressed or enhanced \cite{PhysRevLett.107.107204}. This sort of magnon cooling or heating should also manifest on ultrafast time-scales in either an increase, or a further decrease of the longitudinal magnetization component.

In the following we will discuss simulation results obtained for $\lambda_{\rm {STT}}=2\,$nm at $T=100,\,200,\,300\,$K. In Figure \ref{fig:figure4}(a) we plot the temporal form of the spin current pulse according to Eq.~\eqref{eq:js}. In Fig.~\ref{fig:figure4}(b) we show the temporal evolution of the spatially averaged longitudinal magnetization component $\langle m_x\rangle_{xyz}$ for $T=100\,$K. When the spin current pulse penetrates into the film, the fluctuations increase (decrease), in case of $\mathbf p \downharpoonleft \! \upharpoonright \mathbf x  $ ($\mathbf p \parallel \mathbf x$). Therefore, simultaneously the longitudinal magnetization component further decreases (increases). Note that we only show single time series in Fig.~\ref{fig:figure4}(b). The deterministic nature of the thermal noise in the numerical simulation enables us to apply a normalization procedure to the case when no spins are injected into the ferromagnet. Normalizing and shifting yields the quantity $\langle \Delta m_x\rangle_{xyz}=\frac{\langle m_x\rangle_{xyz}(j_s\neq 0)}{\langle m_x\rangle_{xyz}(j_s=0)}-1 $, which is shown in Fig.~\ref{fig:figure4}(c). This representation clearly shows that the maximum of the change in the magnetic moment is shifted by $360\,$fs with respect to the maximum of the spin current, marked by the vertical dashed line. It can be shown that the front of the transient normalized  $\langle \Delta m_x\rangle_{xyz}$ can be approximated with the time-integrated spin current pulse profile, indicating the accumulative nature of the effect. At later times, both signals decay approximately exponentially with a time constant of $1.79(1)$~ps. Since in our modelling (both numerical and analytical) this decay originates in Gilbert damping only, one can deduce that the spin wave modes with frequencies of a few THz dominate the dynamic response.

We note that the change of the longitudinal moment is with only $10^{-3}$ rather small. As a matter of fact, it increases proportional to the temperature, as Fig.~\ref{fig:figure4}(c) shows. Recall that so far we were discussing spatial averages. By applying the normalization procedure to each layer of the simulated film, we have obtained the spatially resolved data shown in Fig.~\ref{fig:figure4}(d). There one can see that at the injection side the change increases by one order of magnitude, compared to the averaged dynamics. Note that Fig.~\ref{fig:figure4}(d) looks qualitatively similar to the left panel in Fig.~\ref{fig:figure2}(a). However, in contrast to the data shown there, here the spin injection does not cause spatio-temporal coherence of the spin dynamics. Instead, the data plotted in Fig.~\ref{fig:figure4}(c) should be interpreted as a spatio-temporal modulation of the density of thermal magnons in the film.

\begin{figure}
\includegraphics[width=8.25cm]{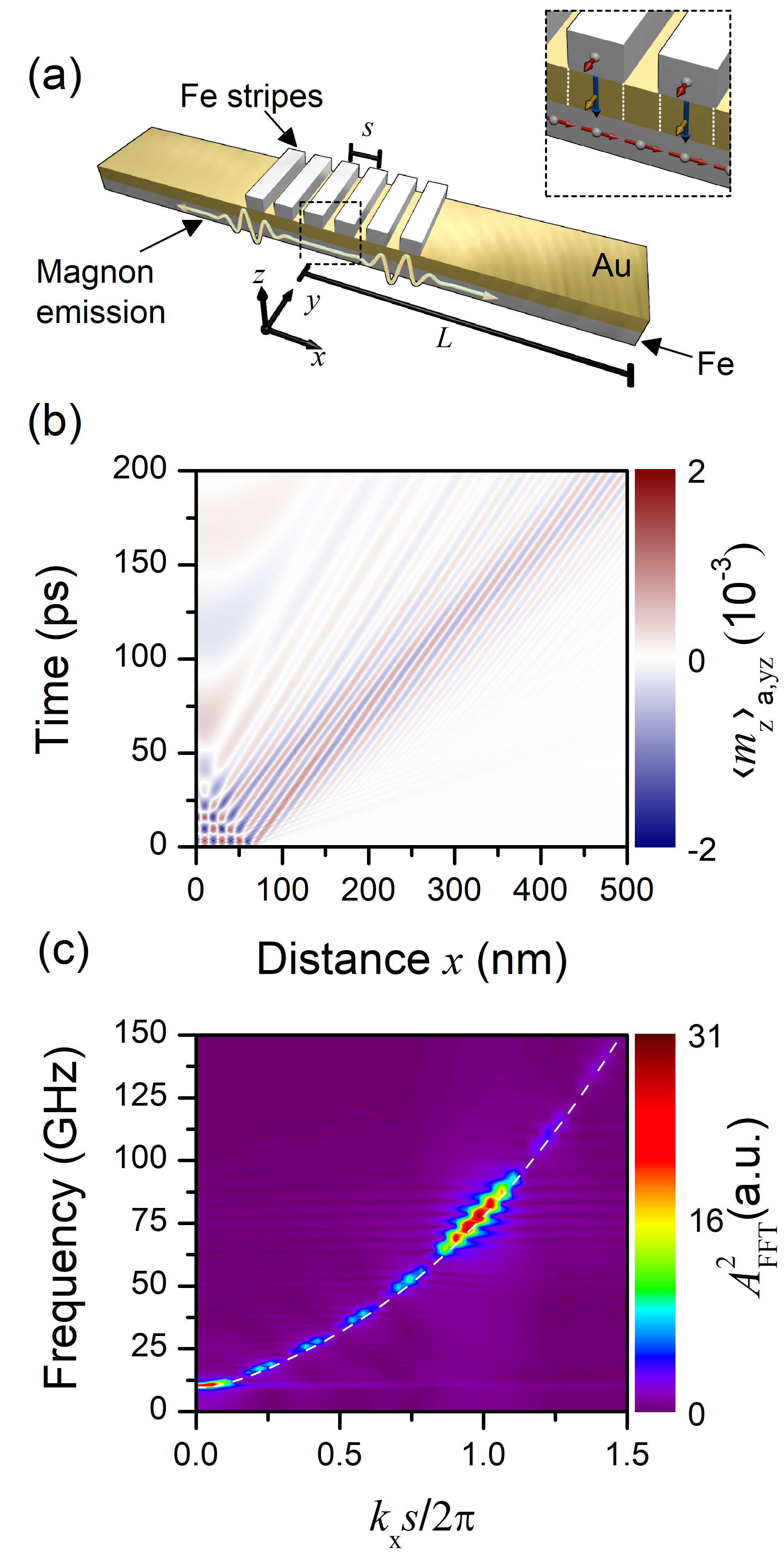}
\caption{Excitation of the in-plane propagating spin waves. (a) Conceptual device design. Femtosecond laser pulse excites the periodically striped top Fe layer, resulting in generation of a spatially-modulated spin current pulse. The latter reaches the underlying orthogonally magnetized Fe layer. The resulted spin transfer torque excites propagating spin waves. Magnifying the active area, the inset shows the spin orientations in the Fe layers, and the spin current flow in the non-magnetic spacer. The stripe period is denoted with $s$, while $L=500\,$nm is the maximum probing length. (b) Spatio-temporal plot of the averaged out-of-plane magnetization component $\langle m_z\rangle_{a,yz}(x,t)$. (c) 2-dimensional Fourier transform of the data shown in (b), indicating efficient excitation of spin waves with wavenumbers governed by the stripes periodicity $s$, i.e. at $k=2\pi/s$. The white dashed line shows the dispersion of dipole-exchange spin waves according to Ref.~\cite{0022-3719-19-35-014}.}
\label{fig:figure5}
\end{figure}

\section{Outlook: Optical spin-wave excitation beyond the diffraction limit}

So far we have been discussing the spin wave (magnon) eigenmodes which are quantized in the direction perpendicular to the film. When considering optical excitation of laterally propagating magnons, one has to acknowledge that this process is usually limited by diffraction. As already mentioned in the introduction, this implies that optical excitation of magnons with wavelengths below the light wavelength
is rather inefficient. We will now show that spin currents offer a unique possibility to overcome this limitation. Consider a multilayer as in Ref.~\cite{Razdolski2017}, but with the top Fe layer patterned into stripes, as depicted in Figure~\ref{fig:figure5}(a). There, the spin current generation, and subsequently the magnon generation in the bottom Fe layer is governed by the geometry of the stripes instead of the laser spot size.  To model this situation, we again employed the spatio-temporal shape of the spin current pulse given by Eq.~\eqref{eq:js}. In addition, a lateral mask defined by six stripes (width $w=10\,$nm and spatial period $s=20\,$nm) was imposed to model the structure shown in Fig.~ \ref{fig:figure5}(a). As such, the lateral cross-section of the laser-generated spin current pulse reproduces the mask stripe pattern. We used $8192\times 1 \times 10$ cells of size $0.5^3\,$nm$^3$, which were enlarged again into a film by applying periodic boundary conditions in $\mathbf x$ and $\mathbf y$-direction.

In Fig.~\ref{fig:figure5}(b) we show a map of the transient magnetization component $\langle m_z\rangle_{a,yz}(x,t)$, averaged across $\mathbf y$, $\mathbf z$, and across a width of $a=2\,$nm around equidistantly probed locations at a distance $x$ towards the center of the stripes. The data clearly show the lateral emission of a spin wave packet, moving with a velocity of about $2500\pm 200\,$m/s.  A Fourier transformation in time and space reveals the spectral properties of the spin dynamics, as shown in Fig.~\ref{fig:figure5}(c). Here, one can see that the dispersion $f(k)$ of dipole-exchange spin-waves in Fe \cite{0022-3719-19-35-014}
(see the dashed line in Fig.~\ref{fig:figure5}(c)) falls on top of the regions of strong response in the spectrum.  Reflecting the periodicity of the stripe pattern, the strongest response can be seen around $f=75\,$GHz, and $k_x=2\pi/s$. The group velocity at this wave number $v_g=\frac{\partial\omega}{\partial k}(2\pi/s)\approx2600\,$m/s is in a good agreement with the propagation speed of the spin-wave packet in Fig.~\ref{fig:figure5}(b).

\section{Conclusions}

In summary, we have shown that the action of ultrafast spin currents penetrating into a magnetic thin film can be modelled with great qualitative and quantitative agreement with the experimental data by including a Slonczewski-like spin-torque term in the micromagnetic equation of motion. Depending on the polarization of the spin current, the resulting torque can either create or annihilate coherent or incoherent magnons. In particular, we reproduced the recent experimental demonstration of coherent magnon generation by spin currents, and obtained further insights into the spatial scales involved in this process. We have identified the factors contributing to the relative excitation efficiency and shown that the linear proportionality law holds over four orders of magnitude. Lastly, employing numerical simulations, we complemented the experimentally estimated constraints on the characteristic spin transfer torque depth $\lambda_{\rm STT}$ in Fe.

Further, our analysis of the spin current excitation of incoherent magnons indicates that the simulated ultrafast heating and cooling should be detectable by magneto-optical methods. We note here that the heating effects which are not explicitly accounted for in our model might introduce additional complications. Both thermal and spin current-driven signals will be overlaid in time, but can in principle be distinguished by their symmetry properties. Considering thinner ferromagnetic films and larger densities of the injected spin current, as in Ref.~\cite{Turgut2013}, our modelling suggests that the hot electrons-driven spin-transfer torque can be a relevant and viable Ansatz to understand the observed spin dynamics. We nevertheless acknowledge that under more extreme conditions a transient change of the magnetization itself, and increased thermal fluctuations cannot be neglected in order to obtain a complete picture. Then, more elaborate modelling techniques such as those described in Ref.~\cite{Evans2012} are needed.

In the outlook section we discussed a way to overcome the diffraction limit when exciting propagating in-plane spin waves. The basic idea relies on the fact in Fe/Au/Fe trilayer spin valve trilayers \cite{Razdolski2017, Alekhin2017}, nanostructuring of the top, laser-excited layer enables spatial tailoring of the spin current profile. In the simplest case, spin currents can only be excited where the top Fe layer exists, thus introducing an in-plane inhomogeneity into the STT stimulus. We have shown that geometrical patterning enables the excitation of propagating spin waves with wavelengths considerably smaller than the optical diffraction limit of typically used VIS to NIR-VIS laser sources. Note that the proposed device design lifts the restriction pertinent to the use of epitaxial Fe films for setting up their magnetizations directions. Indeed, the latter can be achieved exploiting the shape anisotropy of the stripes, and an external magnetic field can be used to ensure an orthogonal magnetization in the bottom Fe layer, instead of relying on magneto-crystalline anisotropy. As such, spin current-driven excitation of high-frequency spin waves in amorphous ferromagnets (e.g. low-loss CoFeB) or even insulating materials (such as yttrium iron garnet attracting increased attention recently) remains an intriguing perspective.

To conclude, we emphasize that this work shows that experimental observations from conventional magnonics and recent ultrafast experiments can be explained on equal theoretical footings. We are convinced that our findings open a fruitful perspective for the application of the predictive and interpretative power of micromagnetic simulation in experimental ultrafast magnetism. 

H.U. acknowledges financial support by the Deutsche Forschungsgemeinschaft within project A06 of the SFB 1073 'Atomic scale control of energy conversion'. The authors thank A.~Melnikov and C.~Seick for valuable comments and M.~Wolf for continuous support.

\section{Supplement}
\subsection{Details of the micromagnetic model}

In Eq.~(1) we neglect the actual propagation of the spin current pulse inside the Fe film with the Fermi velocity $\sim1\,$nm/fs, due to the fact that this speed is much larger than the phase and group velocities of the involved magnons. Furthermore, following theoretical considerations \cite{PhysRevB.70.184438,PhysRevApplied.7.054007}, a field-like torque term was not taken into account in Eq.~(2) due to the condition $d\gg\lambda_{\rm STT}$. 
%
Further experimental support can be obtained from Fig.~4(b) in Ref.~\cite{Razdolski2017}, where the P-MOKE signal proportional to $~m_z$ responds to the STT stimulus with a significant delay. On the contrary, the L-MOKE response immediately follows the spin current-driven accumulation of magnetic moment in Fe, corroborating the dominant role of the damping-like torque term in the STT-induced spin dynamics. If a field-like torque term would be active, the P-MOKE signal $~m_z$ should respond to the spin current excitation directly.

Note that the focus of this work is on accurate modelling of exchange-dominated high-frequency spin dynamics. Given the complicated anisotropy of the spin-wave dispersion \cite{0022-3719-19-35-014} in the dipolar regime, an accurate reproduction of the Ferromagnetic Resonance (FMR) mode and all degenerate in-plane propagating modes is beyond the scope of this work. The reference experiment \cite{Razdolski2017} suggests that the excitation efficiency of this mode is smaller than that for the first-order PSSW (with $n=1$). In our simulations, the experimentally observed ratio of these efficiencies can be achieved by including an additional small external field of $H_{\rm ext}=5\,$Oe, pointing along $\mathbf y$. Variations of the strength of this transverse field affect the FMR mode only while the properties of the PSSW modes remain unchanged. Experimentally, the incidental presence of a transverse magnetic field cannot be excluded either. However, due to uncertainties in its magnitude, we keep $H_{\rm ext}=0\,$, and accept that the excitation efficiency of the FMR mode can be potentially overestimated in our model. 

\subsection{A classical Hamiltonian view}
\subsubsection{General approach}

The theoretical concepts for the following description were first outlined by H. Suhl \cite{SUHL1957209} and V.S. L'vov \cite{lvov1994}. This theory can be regarded as analytic micromagnetic modelling. Note that it has been successfully employed to describe spin-current driven magnetization dynamics in conventional magnonic studies \cite{PhysRevLett.94.037202,Slavin2005,PhysRevLett.95.237201,Demidov2012}. Consider a thin ferromagnetic film of thickness $d=14.2\,$nm, supporting magnons with amplitudes $b_{\mathbf k}$, frequencies $\omega_{\mathbf k}$, and relaxation rates $\Gamma_{\mathbf k}$. According to Suhl, the magnetization dynamics, as described by the Landau-Lifshitz (LL) equation

\begin{equation}
    \mathbf{\dot M}=-\gamma\mu_0 \mathbf M \times \mathbf H_{\rm eff},\label{eq:LL}
\end{equation}

\noindent can be analyzed into plane spin-waves, and the dynamics of these modes can be described by simple kinetic equations. The first step is to linearize the LL-equation \eqref{eq:LL}, and then apply the first Holstein-Primakoff transformation (HPT) to calculate

\begin{eqnarray}
{\dot m^+}=\sum_\mathbf k \dot a_\mathbf k e^{i \mathbf k \mathbf r}={\dot m_y}+i{\dot m_z},
\end{eqnarray}

\noindent where $m_i=M_i/M_0$. The second HPT takes into account ellipticity of the  precession in a tangentially magnetized film. It finally maps $a_\mathbf k$ to the amplitudes $b_\mathbf k$. In total, the HPTs diagonalize the Hamiltonian $\mathcal H$, which generates the LL equation \eqref{eq:LL}. Without dissipation and interactions, $\mathcal H$ then simply reads:

\begin{equation}
    \mathcal H=\sum_{\mathbf k} \hbar\omega_{\mathbf k} b_{\mathbf k}b_{\mathbf k}^*. \label{eq:H}
\end{equation}

Note that for the dispersion $\omega_{\mathbf k}=\omega({\mathbf k})$ we take the approximation by Eq.~(3). The canonical equation of motion for the spin-wave amplitudes is then

\begin{equation}
    {\dot b_{\mathbf k}}+i\omega_{\mathbf k}b_{\mathbf k}=0.
\end{equation}

\noindent Adding a Gilbert-like dissipation term to the Eq.~\eqref{eq:LL}, one gets:

\begin{equation}
   {\dot b_{\mathbf k}}+\left[i\omega_{\mathbf k}+\Gamma_{\mathbf k} \right]b_{\mathbf k}=0,
\end{equation}

\noindent where the relaxation rate is given by

\begin{equation}
  \Gamma_{\mathbf k}=\alpha\omega_H\frac{\partial \omega}{\partial\omega_H},
\end{equation}

\noindent with $\omega_H=\gamma\mu_0 H+\gamma\mu_0\frac{2A}{M_0}k^2$.


\subsubsection{Coherent magnon generation with $\mathbf p\perp \mathbf M$}

For the case of our spatially inhomogeneous spin current, similar to Ref.~\cite{PhysRevLett.94.037202} we first introduce the quantity

\begin{equation}
    \beta(t,\mathbf r)=\gamma\frac{\hbar}{2e \mu_0 M_S} j_0 e^{-z/\lambda_{\rm STT}}\frac{e^{-t/\tau_2}}{1+e^{-(t-t_0)/\tau_1}}.
\end{equation}

\noindent It is then convenient to consider a Fourier representation of $\beta$, and separate out the time-dependence:

\begin{equation}
   \beta(t,\mathbf r)=\theta(t)\sum_\mathbf k \beta_\mathbf k e^{i \mathbf k \mathbf r}.
\end{equation}

\noindent If $\mathbf p\perp \mathbf M$, linearizing the STT term given by Eq.~(2) results for the first HPT in:

\begin{eqnarray}
{\dot m^+}={\dot m_y}+i{\dot m_z}-\theta(t)\sum_\mathbf k \beta_{\mathbf k}e^{i \mathbf k \mathbf r}.
\end{eqnarray}

\noindent The rate equation for a particular mode amplitude $b_\mathbf k$ then reads:

\begin{equation}
{\dot b_{\mathbf k}}+\left[i\omega_{\mathbf k}+\Gamma_{\mathbf k}\right]b_{\mathbf k}-\theta(t) \beta_{\mathbf k}=0.\label{eq:dyn2}
\end{equation}

On short time-scales $t\ll \frac{1}{\Gamma_{\mathbf k}},\,\frac{2\pi}{\omega_{\mathbf k}}$, the second term in Eq.~\eqref{eq:dyn2} can be neglected. Direct integration yields, if the initial dynamic amplitude is small:

\begin{equation}
    b_{\mathbf k}(t)=\beta_\mathbf k\int_0^t \theta(t')dt'.
\end{equation}

\noindent Summation over the all $\mathbf k$ modes and Fourier transformation back into real space gives
 
\begin{equation}
    m_{y}(t,z)=\beta(z)\int_0^t \theta(t^{\prime})dt^{\prime},
\end{equation}

\noindent whereas $m_{z}(t,z)=0$. This result implies that one can obtain the temporal shape of the spin current pulse $\theta(t)$ by taking the time-derivative of $m_y(t)$ probed in the corresponding MOKE geometry (e.g. by L-MOKE in Fig.~4(b) from Ref.~\cite{Razdolski2017}). Furthermore, here it is seen why we call this process coherent. In a stroboscopic pump-probe experiment, one always induces deterministic growth of a transverse magnetization component with the same phase.

For longer time scales, one needs to consider all terms in Eq.~\eqref{eq:dyn2}. Note that a constant spin current $\theta(t)=\theta_0$ leads to a new equilibrium orientation of the magnetization, whereas a time-limited spin current pulse $\theta(t)$ pumps energy into the different spin wave modes. Here we assume that the perturbation is small so that the orientation of the magnetization remains unchanged. Applying a time-domain Fourier transformation to Eq.~\eqref{eq:dyn2} and its complex conjugate we get:

\begin{equation}
\begin{array}{rl}
{i\omega\tilde b_{\mathbf k}}+\left[i\omega_{\mathbf k}+\Gamma_{\mathbf k}\right]\tilde b_{\mathbf k}-\tilde\theta(\omega) \beta_{\mathbf k}=0,\\
i\omega\tilde b_{\mathbf k}^*+\left[-i\omega_{\mathbf k}+\Gamma_{\mathbf k}\right]\tilde b_{\mathbf k}^*-\tilde\theta^*(\omega) \beta^*_{\mathbf k}=0.
\end{array}
\label{eq:dyn3}
\end{equation}

\noindent Thus, the power spectrum of a given mode $\mathbf k$ reads:

\begin{equation}
    p_\mathbf k(\omega)=\tilde b_{\mathbf k}\tilde b_{\mathbf k}^*=\frac{\tilde \theta(\omega)\tilde \theta^*(\omega)\beta_\mathbf k\beta_\mathbf k^*}{\omega_{\mathbf k}^2-\omega^2+\Gamma_{\mathbf k}^2+2i\omega\Gamma_{\mathbf k}}.
\end{equation}

\noindent To obtain the absorbed partial energy $\varepsilon_{\mathbf k}$ introduced in Eq.~(4), we integrate $p_\mathbf k(\omega)$ over the frequency range:

\begin{eqnarray}
    \varepsilon_{\mathbf k}&\sim&{\rm Im}\left[\int_0^\infty p_\mathbf k d\omega\right]\nonumber\\
    &\approx&\tilde \theta(\omega_\mathbf k)\tilde \theta^*(\omega_\mathbf k)\beta_\mathbf k\beta_\mathbf k^*\frac{\tan^{-1}\frac{\Gamma_\mathbf k}{\omega_\mathbf k}}{\omega_\mathbf k}\nonumber\\
    &\approx& \tilde \theta(\omega_\mathbf k)\tilde \theta^*(\omega_\mathbf k)\beta_\mathbf k\beta_\mathbf k^*\frac{\Gamma_\mathbf k}{\omega_\mathbf k^2}.\label{Eq:ppt}
\end{eqnarray}

\noindent Here, we can already identify the factors $|\tilde \theta|^2 (\omega)=\tilde \theta(\omega_\mathbf k)\tilde \theta^*(\omega_\mathbf k)$, and  $|\tilde \beta|^2 (\mathbf k)=\beta_{\mathbf k}\beta_{\mathbf k}^*$ from Eq.~(5). Confirming intuitive expectations, both spectral Fourier powers of the spatial and temporal factors directly impact the partial energy uptake of a particular spin current-driven spin wave mode. Furthermore, we now explicitly derive the additional factor $\xi$ which appears crucial for the comparisons of the excitation efficiency between different modes. Recall that in the numerical simulation, we actually analyze the dynamics of the out-of-plane component $m_z$. When passing back from the amplitudes $b_k$ to the magnetization, one has to take into account the ellipticity $\mathcal{E}_\mathbf k$ of the precession. We thus have to acknowledge that (see e.g. chapter 1 in Ref.~\cite{gurevich1996magnetization})

\begin{equation}
    \frac{\left|m_{\mathbf k,z}\right|^2}{\left|m_{\mathbf k,y}\right|^2}=1-\mathcal{E}_\mathbf k,
\end{equation} 

\noindent where

\begin{equation}
    \mathcal{E}_\mathbf k=\left(1+\frac{\gamma\mu_0 H+\gamma\mu_0\frac{2A}{M_0}k^2}{\gamma\mu_0 M_0}\right)^{-1}.
\end{equation}

\noindent When analyzing the partial energies of the modes found in the spectrum of $m_z$, the mode specific correction yields for the third factor in Eq.~(5):

\begin{equation}
    \xi=\frac{\Gamma_\mathbf k}{\omega_\mathbf k^2}\frac{1-\mathcal{E}_\mathbf k}{2-\mathcal{E}_\mathbf k}.
\end{equation}


\subsubsection{Incoherent magnon generation with $\mathbf p\parallel \mathbf M$}

We will now consider spin injection into a thermally occupied magnon ensemble. If we assume a spatially homogeneous spin current, whose polarization $\mathbf p\parallel \mathbf M$, the linearized contribution of the STT term to the LL-equation \eqref{eq:LL} results in:

\begin{eqnarray}
{\dot m^+}={\dot m_y}+i{\dot m_z}+\beta(t)m^+,
\end{eqnarray}

\noindent with the temporal dependence of $\beta$ given by Eq.~(1). The rate equation for $b_{\mathbf k}$ reads \cite{PhysRevLett.94.037202}:

\begin{equation}
    {\dot b_{\mathbf k}}+\left[i\omega_{\mathbf k}+\Gamma_{\mathbf k}+\beta(t) \right]b_{\mathbf k}=\mathcal F_\mathbf k,\label{eq:dyn}
\end{equation}

\noindent where $\beta(t)=\gamma\frac{j_s(t)}{d\mu_0 M_0}$, and $\mathcal F$ represents a thermal noise source. Note that in Eq.~\eqref{eq:dyn} higher order terms in $b_\mathbf k$, which result from the second HPT, are not taken into account \cite{PhysRevLett.94.037202}. Note that the product $\beta(t)b_\mathbf k$ in Eq. \eqref{eq:dyn} explains why the excitation process is incoherent: The random phase of a thermally driven magnon gets imprinted on the spin current term. Therefore in a pump-probe experiment, the temporal evolution of the individual mode's phases will differ from shot to shot in a random fashion. 

If the spin current is spatially inhomogeneous, one should again consider $\beta=\theta(t)\sum_\mathbf k \beta_\mathbf k e^{i \mathbf k \mathbf r} $. Then, in principle, different $\beta_\mathbf k$ induce a mixing between the magnon modes $b_\mathbf k$. The rate equation then reads:

\begin{eqnarray}
    {\dot b_{\mathbf k}}&+&\left[i\omega_{\mathbf k}+\Gamma_{\mathbf k}\right]b_{\mathbf k}\nonumber\\
    &+&\sum_{\mathbf{k'},\mathbf{k''}}\delta(\mathbf{k'}+\mathbf{k''}-\mathbf k)\beta_\mathbf{k'}b_\mathbf{k''}=\mathcal F_\mathbf k.\label{eq:dyn}
\end{eqnarray}

\noindent For the experimental situation of a laterally homogeneous, but vertically inhomogeneous spin current, the mixing couples the modes with different $k_z$ but equal $k_x$ and $k_y$. To simulate the dynamics of a thermal magnon ensemble, we consider a volume of $v=dL^2$, and quantize the wavenumbers according to $k_x=\frac{n\pi}{L}$, $k_y=\frac{m\pi}{L}$, $k_z=\frac{o\pi}{d}$, with $L=10d$. Modes up to $m,n=\pm 256$, and $o=24$ are taken into account. For the implementation of $\mathcal F_\mathbf k$ we have chosen a Gaussian random number generator, which obeys

\begin{equation}
    \langle\mathcal F_{\mathbf k} F_{\mathbf k'}^* \rangle =2\delta(\mathbf k-\mathbf k')\Gamma_{ \mathbf k} \frac{k_{\rm B} T}{\hbar\omega_\mathbf k}.
\end{equation}

\noindent In the absence of a spin current, this provides in the temporal average an equilibrium magnon number density of

\begin{equation}
        |b_\mathbf k^{eq}|^2=\langle b_\mathbf k b_\mathbf k^*\rangle=\frac{k_{\rm B} T}{\hbar\omega_\mathbf k}.
\end{equation}

\noindent To probe the dynamics, we determine the total number of magnons $\sum_\mathbf k b_\mathbf k b_\mathbf k^* $. Since each magnon carries $2\mu_B$, this reduces the magnetization to 

\begin{equation}
    m_x=\sqrt{1-\frac{2\mu_B}{v M_s}\sum_\mathbf k b_\mathbf k b_\mathbf k^*}
\end{equation}

\noindent Similar to that in the analysis of the numerical simulations output, we then normalize the results by the equilibrium magnetization $m_x^{\rm eq}$

\begin{equation}
    m_x^{\rm eq}=\sqrt{1-\frac{2\mu_B}{v M_s}\sum_\mathbf k \frac{k_{\rm B} T}{\hbar\omega_{\mathbf k}}},
\end{equation}

\noindent and shift by $1$ to obtain 

\begin{equation}
    \Delta m_x=\frac{m_x}{m_x^{\rm eq}}-1.\label{eq:dmz}
\end{equation}

\begin{figure}[h!]
\includegraphics[width=8.25cm]{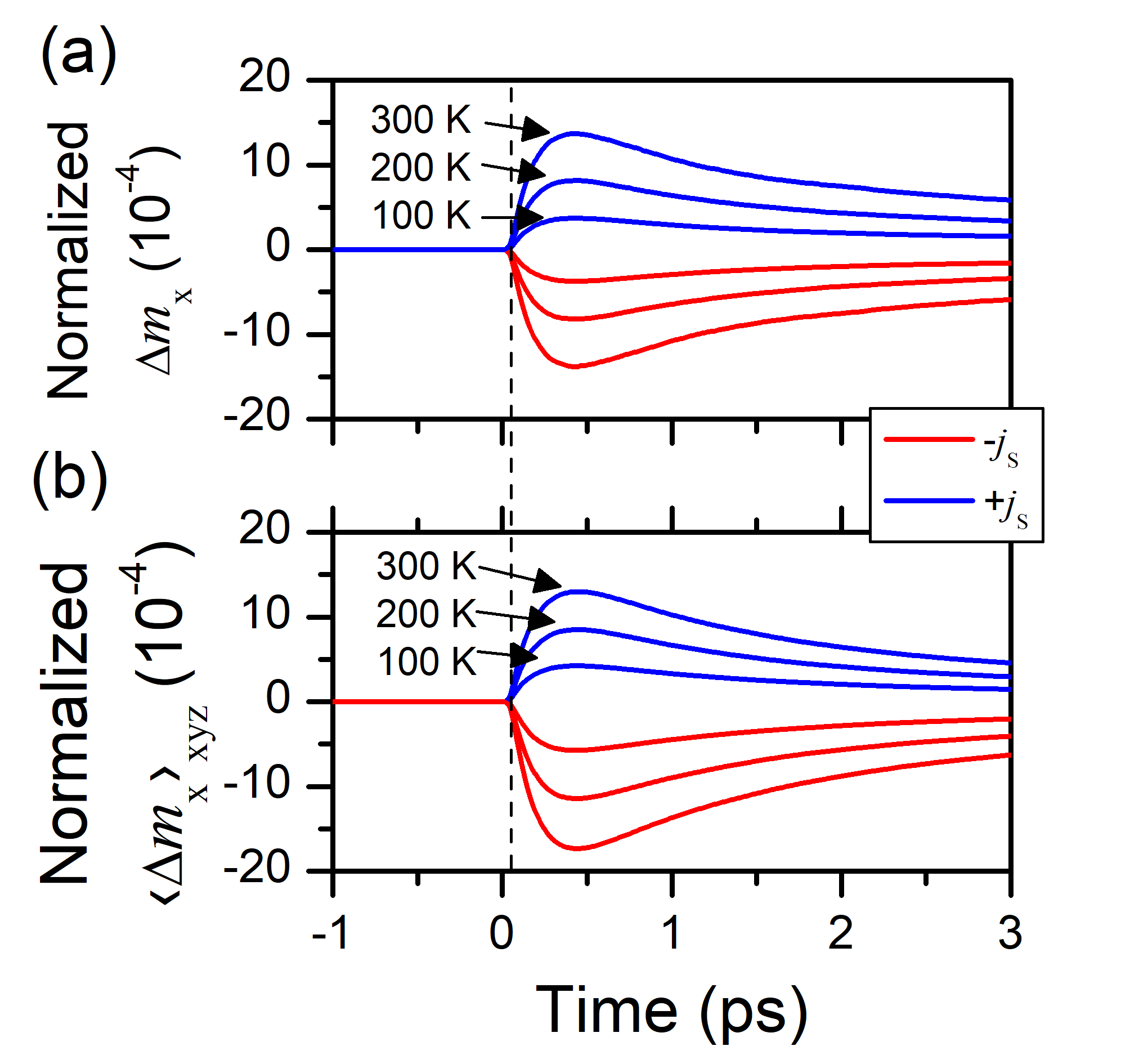}
\caption{Comparison of the spin dynamics according to the classical Hamiltonian model and to the numerical micromagnetic modelling. The red and blue lines are obtained for the opposite polarities of the spin current pulse. (a) Relative variations of averaged magnon number $\langle\Delta m_x\rangle$ from Eq.~\eqref{eq:dmz} as a function of time, obtained from the Hamiltonian model given by Eq.~\eqref{eq:dyn}. Computation for different temperatures as indicated. (b) Corresponding micromagnetic simulation data, also shown in Figure 6(c). The dashed line indicates the peak time of the injected spin current pulse.}
\label{fig:figureS1}
\end{figure}

Note that for comparing this analytic model with the micromagnetic model, we neglected the cross coupling terms in Eq.~\eqref{eq:dyn} since those cancel out in the statistical average. Figure \ref{fig:figureS1} shows the results obtained for positive and negative spin current pulses of equal magnitude, at three different temperatures. The data look quite similar to those shown in Figure 6(c). For instance, for positive spin injection, after a fast initial increase which saturates $350\,$fs after the maximum of the injection density, the magnon numbers decrease approximately exponentially with a decay time of $1.48(1)\,$ps. This value is quite close to the one obtained in the numerical simulations ($1.79(1)\,$ps). Small differences in decay time and in the overall magnitude can be attributed to the fact that in the rate equations \eqref{eq:dyn} interactions between the magnons are not taken into account. In contrast, in the micromagnetic simulation, the underlying fully nonlinear Landau-Lifshitz equation captures these processes. Also, slightly different compositions of the magnon ensembles in the analytic and numerical description can give rise to different decay rates.

\bibliography{source}

\end{document}